\documentclass[12pt,times,tight]{aastex62}

\usepackage{amsfonts}
\usepackage{mathrsfs}
\usepackage{bm}
\usepackage{lipsum}
\usepackage{physics}
\usepackage{multirow}
\usepackage{rotating}
\usepackage{graphicx}

\def\bmell{{\bm{\ell}}}
\def\bhm{M_{\bullet}}
\def\bmalpha{{\bm{\alpha}}}
\def\bmn{\bm{n}_{\rm obs}}
\def\calA{{\cal A}}

\def\calO{{\cal O}}
\def\cblue{\color{black}}

\def\fline{f_{\ell}}
\def\Fline{F_{\ell}}
\def\sunm{M_{\odot}}

\received{\today}
\revised{***}
\accepted{***}

\shorttitle{Signals of Binary Supermassive Black Holes}
\shortauthors{Songsheng et al. }

\begin{document}

\title{\bf \large
Differential Interferometric Signatures of Close Binaries of Supermassive Black Holes in \\
Active Galactic Nuclei}

\correspondingauthor{Jian-Ming Wang}
\email{wangjm@ihep.ac.cn}

\author{Yu-Yang Songsheng}
\affil{Key Laboratory for Particle Astrophysics,
Institute of High Energy Physics, Chinese Academy of Sciences,
19B Yuquan Road, Beijing 100049, China}
\affil{University of Chinese Academy of Sciences,
19A Yuquan Road, Beijing 100049, China}

\author{Jian-Min Wang}
\affil{Key Laboratory for Particle Astrophysics,
Institute of High Energy Physics, Chinese Academy of Sciences,
19B Yuquan Road, Beijing 100049, China}
\affil{University of Chinese Academy of Sciences,
19A Yuquan Road, Beijing 100049, China}
\affil{National Astronomical Observatories of China,
Chinese Academy of Sciences,
20A Datun Road, Beijing 100020, China}

\author{Yan-Rong Li}
\author{Pu Du}
\affil{Key Laboratory for Particle Astrophysics,
Institute of High Energy Physics, Chinese Academy of Sciences,
19B Yuquan Road, Beijing 100049, China}

\affil{University of Chinese Academy of Sciences,
19A Yuquan Road, Beijing 100049, China}

\begin{abstract}
In the present paper, we explore opportunities of applying the GRAVITY at the 
Very Large Telescope Interferometry (VLTI) with unprecedented 
spatial resolution to identify close binaries of supermassive black holes
(CB-SMBHs) in active galactic nuclei (AGNs). Each SMBH is assumed to be separately 
surrounded by their own broad-line regions (BLRs) composed of clouds with virialized
motion. Composition of the binary orbital motion 
and the virial motion of clouds in each BLR determines the projected 
velocity fields and hence differential phase curves, which are obviously different 
from that of a single BLR. We calculate emission line profiles and differential 
phase curves of CB-SMBHs for the GRAVITY. For the simplest case where angular 
momentums of two BLRs and orbital motion are parallel, a phase plateau generally 
appears in the phase curves. For other combinations of the angular momentum, the
plateau is replaced by new peaks and valleys variously depending
on the situations. Given a combination, phase curves are also sensitive to 
changes of parameters of CB-SMBHs. All these
features are easily distinguished from the well-known $S$-shaped phase curves of a 
single BLR so that the GRAVITY is expected to reveal signals of CB-SMBH from candidates 
of AGNs. With joint analysis of observations of reverberation mapping campaigns, we 
can reliably identify CB-SMBHs, and measure their orbital parameters in the meanwhile.
This independent measurement of the orbital parameters also has 
implications to analysis of Pulsar Timing Array (PTA) observations for
properties of low-frequency gravitational waves in future.
\end{abstract}


\section{Introduction}
In the framework of hierachical model of galaxy formation and evolution, binary supermassive 
black holes (SMBHs) bounded by their gravity are a natural consequence of evolution of galaxies 
and black holes from dual active galactic nuclei 
(AGN) \citep[e.g.,][]{Komossa2003,Comerford2009,Wang2009,Fu2011,Ge2012,Liu2018} to close binaries 
of SMBHs (CB-SMBHs) with separations less than 0.1pc ($\sim 120$ltd) 
and finally merge \citep{Begelman1980}. Given a CB-SMBH with the total mass and the separations, 
its orbital period is
\begin{equation}
{\cal T}_{\rm orb}=227.4\,M_{8}^{-1/2}\calA_{100}^{3/2}\,{\rm yrs},
\end{equation}
where $M_{8}=M_{\rm tot}/10^{8}\sunm$ is the total mass and $\calA_{100}=\calA_{0}/100{\rm ltd}$ 
is the separation of the binary in units of 100\,ltd.
Low-frequency gravitational waves radiated from the CB-SMBH in the Universe
have been explored for more than 10 years by several long-term campaigns of Pulsar Timing 
Array (PTA) \citep[see a review of][]{Manchester2013,Lommen2015,Mingarelli2019}. However, detections 
show that the characteristic amplitude of GW background is less than $1.0\times 10^{-15}$ 
with 95\% confidence \citep[e.g.][]{Shannon2015,Mingarelli2019}, which is under the predictions 
from the SMBH binary synthesis \citep[e.g.,][]{Mingarelli2017}, or individual 
CB-SMBH \citep{Arzoumanian2014,Wang2017,Sesana2018,Aggarwal2018,Hobbs2019}. 
CB-SMBHs are of great interests to astronomers and 
physicists for low-frequency gravitational waves, and evolution of galaxies and black holes,
but they are observationally elusive completely. Searching for SMBH binaries is urgent, and it 
is more important to measure orbital parameters of the CB-SMBHs from independent ways to test 
low-frequency GW detected PTA observations {\cblue since the binary parameters could be poorly
estimated only by the PTA \citep[see an extensive review of][]{Burke-Spolaor2019}}. 

CB-SMBHs are so close that it is difficult to image using current optical technology. 
Complicated profiles of broad emission lines as an indicator in quasars
was originally suggested by \cite{Gaskell1983}, but they could be alternatively 
explained by other models, such as precessing spiral arm \citep{Eracleous1995,Storchi-Bergmann2003} 
or hot spot \citep{Newman1997,Jovanovic2010} (see \S\ref{sec:alter} for a brief discussion).
Except for some features of optical emission lines \citep[see a review of][]{Popovic2012},
there are several promising ways to search for them from: 1) periodicity of long-term photometric 
variations of AGNs \citep{Graham2015,Charisi2016}, in particular, OJ 287 with a period of 
$\sim 11$yrs \citep{Valtonen2008}\footnote{Actually it is quite hard to determine the periodicity
of long term variations since it usually needs at least length of monitoring campaigns to be
three or four cycles of period to rule out pseudo-periodicity \citep[e.g.,][]{Li2019}. 
Unless AGN targets have shorter periods, campaigns of 
long-term variations are very time consuming.}; 2) long-term variations of broad emission line 
profiles (e.g., the radial velocity curves) \citep[e.g.,][]{Shen2010,Pflueger2018}, such as in 
NGC 4151 \citep{Bon2012} and NGC 5548 \citep{Li2016}, and a large size of sample
to be carried out for long-term profile variations \citep[e.g.,][]{Runnoe2017,Guo2019}; 
3) AGNs with X-shaped jets \citep{Merritt2002,Kharb2017}; 4) deficit of ultraviolet 
continuum \citep{Gultekin2012,Yan2015}; 5) 2-dimensional transfer functions obtained by 
reverberation mapping of AGNs, depending on the orbital motion of binary SMBHs 
\citep{Wang2018,Songsheng2019}. A long-term project of 
Monitoring AGNs with H$\beta$ Asymmetry (MAHA) is being conducted through the Wyoming Infrared 
Observatory (WIRO) 2.3m telescope for a systematical search for CB-SMBHs in the local 
Universe \citep{Du2018,Brotherton2019} and show preliminary evidence of 2D transfer functions 
for appearance of CB-SMBHs in Mrk 6 and Akn 120 from $\sim 50$ targets. In principle,
kinematics of ionized gas can be mapped through monitoring reverberations of broad emission 
lines to distinguish binary black holes from a single black hole \citep{Wang2018}.
More campaigns are expected for appearance of CB-SMBHs.

Thanks are given to the GRAVITY at the Very Large Telescope Interferometry (VLTI)
with its unprecedented spatial resolution ($\sim 10\mu$as) \citep{Abuter2017} for a successful
detection revealing a disk-like structure of broad-line region in 3C 273 \citep{Sturm2018}.
The GRAVITY results are in excellent agreement with that from a ten-year campaign of 
reverberation mapping (RM) of 3C 273 with homogeneous and high cadence \citep{Zhang2019}. This 
demonstrates the powerful and efficient capability of the GRAVITY spatial resolution.
In principle, the GRAVITY and RM-campaigns can provide independent measurements (actually
supplementary to each other), which obtain information of the ionized gas in directions 
perpendicular and parallel to the line of observer's sight, respectively. 
In this paper, we apply the interferometry technique to search for signals of CB-SMBHs. 
We demonstrate that the composite velocity of ionized gas in galactic centers
governs $\lambda$-photoncenters of the binary system, showing 
distinguished differential phase curves from the well-known $S$-shaped curves in a single BLR. 
This allows us to extract part 
of spatial information of non-resolved sources using optical interferometry of the VLTI 
for candidates of low-frequency gravitational waves. 

The paper is scheduled as the following. We present the fundamental formulations of the 
interferometry in \S2. The differential phase curves are provided in \S3 and \S4 for 
a single BLR and binary BLRs, respectively, illustrating unique features of CB-SMBHs. 
Discussions on the issues of binary black holes are given in \S5.  
We draw conclusions in the last section. 

\begin{figure}\label{fig:coordinate}
    \centering
    \includegraphics[scale = 0.7]{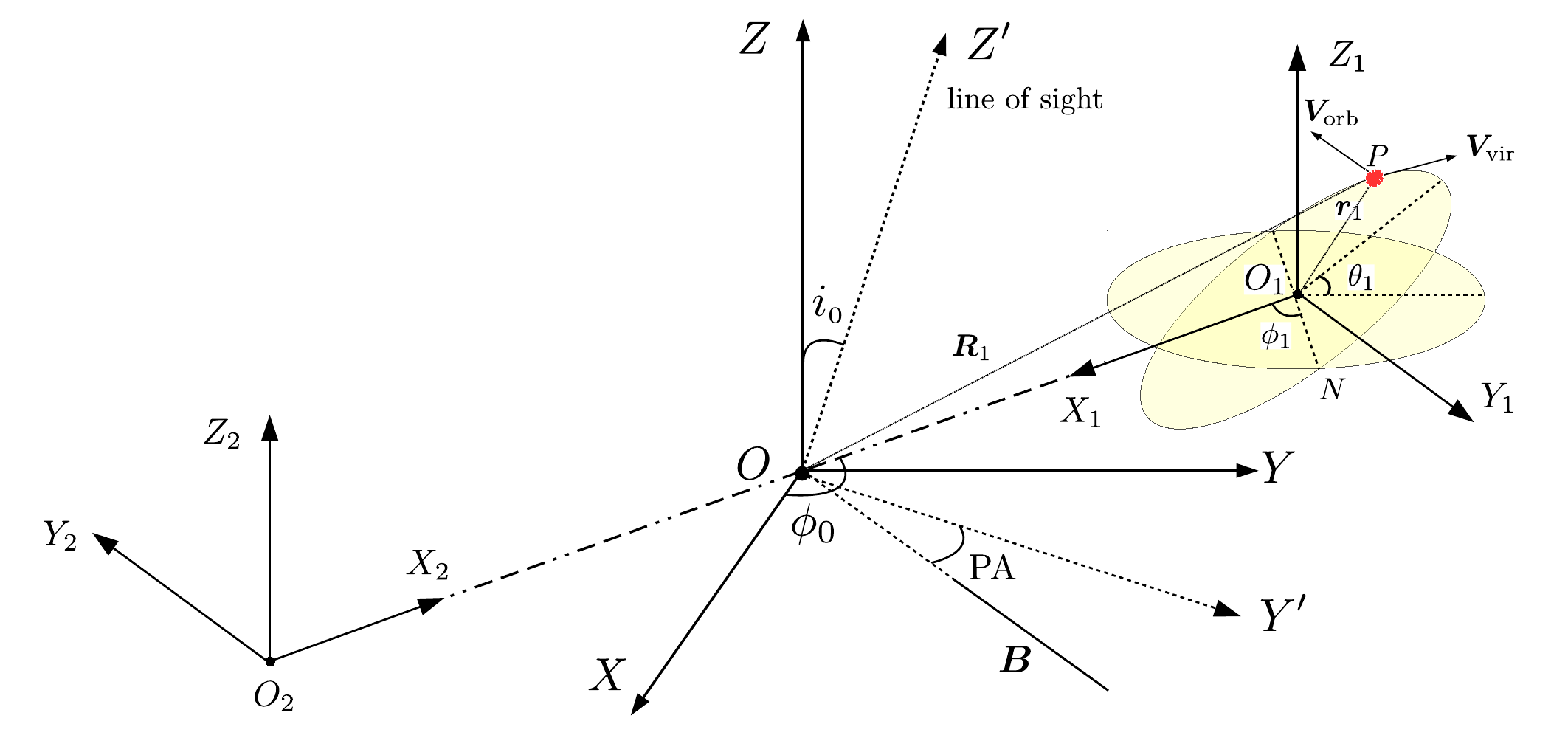}
    \caption{Coordinate system used in the present model. $O$ is the mass center of 
    the binary system and $O-XY$ is its orbital plane. A remote observer 
    with an inclination of $i$ to the $OZ$-axis is
    located in the $O-YZ$ plane. $OY^{\prime}$ is perpendicular to the $O-XZ^{\prime}$ 
    plane. The $O-XY^{\prime}$ plane is
    the tangent plane of the sky since it is perpendicular to the LOS of the observer.  
    $OZ^{\prime}$ is the LOS.
    $O_{1}$ and $O_{2}$ are the primary and the secondary centers, respectively. 
    For simplicity, we only plot the primary BLR and omit the secondary.
    In the binary system, $O_{1}Z_{1}\parallel O_{2}Z_{2}\parallel OZ$ holds and the binary BLRs 
    are co-planed with the orbit. $\phi_{0}$ is the phase angle of orbital motion starting from $OX$-axis.
    $\theta_{1}$ and $(\phi_{1}-\pi/2)$ are the polar and azimuth angles 
    of the normal line of one cloud's orbital plane in BLR-I, respectively ($\theta_{1}$ and $\phi_{1}$ 
    are corresponding angles in BLR-II). One cloud (red point) in BLR-I is located at $P$ with virial 
    velocity $\bm{V}_{\rm vir}$ additional to the orbital motion ($\bm{V}_{\rm orb}$) in the observer's 
    frame. $\phi_{\rm B1} = \angle NO_1P$ is the orbital phase angle of that cloud. All the parameters
    and the angles of the BLR-II can be easily obtained by comparing with BLR-I, we thus
    omitted in this plot. The baseline $\bm{B}$ is in the $O-XY^{\prime}$ plane with a position angle
    (PA).}
\end{figure}

\section{Optical interferometry of BLRs}
\subsection{Basic formulations}
Interferometry provides high spatial resolution of compact objects through differential phases measured by
a pair of telescopes separated by a distance. The phases indicate the path difference of photons emitted 
from a point to different telescopes, allowing astronomers to get spatial information of the objects.  
We follow the description of interferometry given by \cite{Petrov1989}, but see also \cite{Rakshit2015}.
For an interferometer with baseline $\bm{B}$, a non-resolved source with a global angular size smaller than 
the interferometer resolution limit $\lambda / B$ has the interferometric phase given by
\begin{equation}\label{eq:phase}
\phi_*(\lambda,\lambda_{\rm r}) = -2\pi\bm{u} \vdot [\bm{\epsilon}(\lambda) - \bm{\epsilon}(\lambda_{\rm r})],
\end{equation}
where
\begin{equation}\label{eq:ph-center}
\bm{\epsilon}(\lambda) = 
\frac{\iint \bm{\alpha} \calO(\bm{\alpha},\lambda)  \dd[2]{\bm{\alpha}}}{\iint \calO(\bm{\alpha},\lambda) \dd[2]{\bm{\alpha}}} \qc \bm{u} = \frac{\bm{B}}{\lambda},
\end{equation}
$\calO(\bm{\alpha},\lambda)$ is the surface brightness distribution of the source,
$\bm{\alpha}$ is the angular displacement on the celestial sphere,
and $\lambda_{\rm r}$ is the wavelength of a reference channel. 
Here $\bm{\epsilon}$ is the photoncentre of the source at wavelength $\lambda$ and $\bm{u}$ 
is the spatial frequency. Here we point out that the
bold letters are vectors, and stress that all the expressions in \S2.1 and \S2.2 are in a general 
form regardless of coordinate frames.

For a single AGN, the continuum emission in the $K$-band mainly originates from the hot dust near 
the sublimation radius \citep{Kishimoto2007, Kishimoto2009}, while the emission lines are generated 
by the photon-ionized gas in the BLR. We have 
the total surface brightness distribution given by
\begin{equation}
\calO(\bm{\alpha},\lambda) = \calO_{\rm dust}(\bmalpha,\lambda) + \calO_{\rm BLR}(\bm{\alpha},\lambda),
\end{equation}
where $\calO_{\rm dust}(\bmalpha,\lambda)$ is surface brightness of the dust torus, and 
$\calO_{\rm BLR}(\bm{\alpha},\lambda)$ is that of the BLR. For simplicity, we choose to model 
the dusty torus as a uniform ring with photocenter located at the black hole, i.e.,
\begin{equation}\label{eq:s-bright}
\bm{\alpha}_{\bullet}=\frac{\iint\bm{\alpha}\calO_{\rm dust}(\bmalpha,\lambda)\dd[2]{\bm{\alpha}}}
                       {\iint\calO_{\rm dust}(\bmalpha,\lambda)\dd[2]{\bm{\alpha}}},
\end{equation}
where $\bm{\alpha}_{\bullet}$ is the coordinate of the central black hole on the celestial sphere.
%
%
%
Given geometry and kinematics of a BLR, its $\calO_{\rm BLR}(\bm{\alpha},\lambda)$ can be calculated 
for one broad emission line with the observed central wavelength $\lambda_{\rm cen}$ thorough
\begin{equation}
\calO_{\rm BLR}(\bm{\alpha},\lambda) = 
\int \frac{\xi(\bm{r})F_{\rm c}}{4 \pi r^2} f(\bm{r},\bm{V}) \delta\!\left[\bm{\alpha} 
       - \bmalpha_{\bullet} - \frac{\bm{r} - 
(\bm{r}\vdot\bmn)\bmn}{D_{\rm A}}\right] 
\delta\!\left[\lambda - \left(1 + \frac{\bm{V}\vdot\bmn}{c}\right) \lambda_{\rm cen}\right] 
         \dd[3]{\bm{r}} \dd[3]{\bm{V}},
\end{equation}
where $\bm{r}$ is the displacement to the central BH, 
$\xi(\bm{r})$ is the reprocessing coefficient at position $\bm{r}$ and $f(\bm{r},\bm{V})$ 
is the velocity distribution at that point, $F_{\rm c}=L_{\rm c}/4\pi D_{\rm L}^{2}$ is fluxes
received by an observer,
$L_{\rm c}$ is the luminosity of ionization continuum radiation from accretion disk,
$D_{\rm L}$ and $D_{\rm A}$ are the luminosity and the angular size distance of the AGN
and $\bmn$ is the unit vector pointing from the observer to the source \citep{Blandford1982}.
For simplicity, we choose $\bmalpha_{\bullet} = 0$, and have 
\begin{equation}\label{eq:epsilon}
\bm{\epsilon}(\lambda) = \bm{\epsilon}_{_{\rm BLR}}(\lambda) \fline(\lambda),
\end{equation}
where
\begin{equation}
\bm{\epsilon}_{_{\rm BLR}}(\lambda) = \frac{\int\!\!\!\int \bm{r} \calO_{\rm BLR}(\bm{\alpha},\lambda)  
\dd[2]{\bm{\alpha}}}{\int\!\!\!\int \calO_{\rm BLR}(\bm{\alpha},\lambda) \dd[2]{\bm{\alpha}}},\, 
\fline(\lambda) = \frac{\Fline(\lambda)}{\Fline(\lambda) + F_{\rm c}(\lambda)},\, 
\Fline(\lambda) =\!\! \int\!\!\!\!\int\!\! \calO_{\rm BLR}(\bm{\alpha},\lambda) \dd[2]{\bm{\alpha}},\,
F_{\rm c}(\lambda)=\!\!\int\!\!\!\!\int\!\! \calO_{\rm dust}(\bmalpha,\lambda)\dd[2]{\bm{\alpha}},
\end{equation}
by inserting Equation (\ref{eq:s-bright}) into (\ref{eq:ph-center}). The phase curves can be
determined by Eqs (\ref{eq:phase}) and (\ref{eq:epsilon}). This technique has been successfully
applied to 3C 273 \citep{Sturm2018}, revealing a disk-like BLR emitting Paschen $\alpha$ in this quasar.
The phase curves can be used to probe dynamics and structure of ionized gas in AGNs.

\subsection{Interferometry for binary BLRs}
We are applying this technique to search for signals of binary black holes. This can be done by
taking $\calO(\bmalpha,\lambda)$ composed of two separated sources. The binary system has
the primary and the secondary masses of $M_1$ and $M_2$ with an separation of $d_{\bullet}$, 
respectively. In the case of binary BLRs, we choose the mass 
center of the CB-SMBHs as origin of the coordinates. As a result, we
%
have (see Appendix for derivation)
\begin{equation}\label{eq:alpha}
\bm{\alpha}^{\bullet}_{1} = \frac{\bm{\calA}_{0} - (\bm{\calA}_{0}\vdot\bmn)\bmn}{D_{\rm A}}\,(1-\mu_1)\qc 
\bm{\alpha}^{\bullet}_{2} = -\frac{\bm{\calA}_{0} - (\bm{\calA}_{0}\vdot\bmn)\bmn}{D_{\rm A}}\,\mu_1,
\end{equation}
for each SMBHs, where $\bm{\calA}_{0}$ is the displacement of the secondary BH relative to the 
primary BH, and $\mu_1=M_1/(M_1+M_2)$ is the fraction of the primary BH to the total mass 
of the system. The surface brightness distribution should be generalized to
\begin{equation}\label{eq:total-surface}
\calO_{\rm I+II}(\bm{\alpha},\lambda) = \calO_{\rm dust}(\bm{\alpha},\lambda)
              +\sum_{i=1}^2\calO_{{\rm BLR},i}(\bm{\alpha},\lambda).
\end{equation}
{\cblue In our modeling of CB-SMBHs, the separations are around three times the average 
radius of the BLR of the primary BH. Since the dust reverberation radius is systematically larger 
than that of the broad Balmer emission lines by about a factor of four 
to five \citep{Koshida2014}, we would assume the binary BLRs are encompassed by a common dust 
torus. The photoncenter of the torus is hard to determine. However, the $\lambda$-photoncenter 
of the continuum can be approximated as fixed in the wavelength range of a specific emission line.
%
So the photoncenter location can only lead to vertical shifts of the 
differential phase curve. In order to illustrate the shape of the differential phase curve of binary 
BLRs, we would simply choose the location of photoncenter to be the mass canter of the system.}

Given geometry and kinematics of binary BLRs, the interferometric phase can then be calculated 
by Equations (\ref{eq:phase}) and (\ref{eq:ph-center}) straightforward
\begin{equation}\label{eq:calOIII}
\calO_{\rm BLR}^{\rm I+II}(\bm{\alpha},\lambda) = \sum_{i=1}^{2}
   \int \frac{\xi_i(\bm{r}_i)F_{i}^{\rm c}}{4\pi r_{i}^2} f_{i}(\bm{r}_{i},\bm{V}_{i}) 
    \delta\!\left[\bm{\alpha} 
   - \bmalpha^{\bullet}_{i} - \frac{\bm{r}_{i} - (\bm{r}_{i}\vdot\bmn)\bmn}{D_{\rm A}}\right] 
   \delta\!\left[\lambda - \left(1 + \frac{\bm{V}_{i}\vdot\bmn}{c}\right)\lambda_0\right] 
    \dd[3]{\bm{r}}_{i} \dd[3]{\bm{V}}_{i}.
\end{equation}
Substituting Eq. (\ref{eq:alpha}), (\ref{eq:total-surface}) and (\ref{eq:calOIII}) 
into (\ref{eq:phase}) and (\ref{eq:ph-center}), we have the differential phase curves for binary BLRs. 
From Eq. (\ref{eq:calOIII}),
we know two aspects controlling $\calO_{\rm BLR}^{\rm I+II}$ through the projected distance (the first
bracket) and projected velocity fields (the second bracket). The first is determined by the relative 
orientations of the orbital plane and the LOS of observer. The second is the total velocity fields 
composited of orbital motion and local virial motion of clouds in each BLR. It should
be noted that these equations apply to any kinds of geometry and kinematics. To capture the major physics
of the complicated system, in the following sections, we assume the simplest model of the binary 
BLRs for their distinguished features in the following sections.

\subsection{Geometry and kinematics}
Coordinates of one CB-SMBH are shown in Figure \ref{fig:coordinate}. 
Each BH has its own BLR detached from each other. For the simplest version of the present model, we assume 
that individual BLR is of disk-like structure with axial symmetry by neglecting interaction of another
BH. This can be justified by velocity-resolved delay maps obtained by reverberation mapping 
of AGNs. Disk-like geometry of the BLR is favored in most AGNs, such as, 
NGC 5548 \citep[e.g.,][]{Grier2015,Bentz2013,Lu2016,Du2016,Xiao2018} though outflows and inflows 
appear in a few AGNs. Particularly, a disk-like structure of the BLR in 3C 273 is favored by 
GRAVITY observations in \cite{Sturm2018}.
Each BLR with axis-symmetry around its BH can be described by inner and outer radii ($r_{\rm in}$ and
$r_{\rm out}$, respectively), and its half opening angle ($\Theta$). In this paper, we also assume that 
the two BLRs and the binary orbit are co-planed in the $O-XY$ plane for the simplest case. It is possible 
that the binary black holes and two BLRs are not co-planed. This will be
treated in another future paper.

We also assume the reprocessing coefficient of the clouds in each BLR decays radially as a power-law 
\begin{equation}
\xi(r) = \xi_0 \left(\frac{r}{r_{\rm in}}\right)^{-\gamma},
\end{equation}
with the power index $\gamma$ and normalization $\xi_0$. This distribution 
is supported in the BLR modelling through RM data \citep{Pancoast2011,Li2013}. 

For simplicity, we assume that the binary black holes are rotating around 
their mass center in a circle orbit with the Keplerian angular velocity
of $\Omega=\sqrt{G(M_1+M_2)/\calA_{0}^{3}}$, where $G$ is the gravitational 
constant. Clouds in each BLR are virilaized with fully ordered motion, 
namely, directions of their angular momentum are alway upward or downward 
simultaneously. Virialized motion in a few AGNs has been tested through
reverberation mapping campaigns. It has been found that $V_{\rm FWHM}\propto \tau_{\rm H\beta}^{-2}$
holds in NGC 5548, 3C 390.3 and NGC 7469 from multiple campaigns \citep{Peterson2000}, 
where $\tau_{\rm H\beta}$ is the H$\beta$ lags with respect to the varying continuum 
and $V_{\rm FWHM}$ is the full-wdith-half-maximum of H$\beta$ profile. Moreover, the
degree of ordered motions of clouds in the BLR has influence on the differential phase
curves \citep{Stern2015}, however, it is very hard to justify this situation. 
Fully ordered motion of clouds as an important assumption has been used in \cite{Sturm2018}.
{\cblue In this paper, we assume fully ordered motion of clouds in each BLRs. This makes the phase
signal to be maximized when fixing other parameters.}
\cite{Stern2015} had noted this effects, 
but this paper pay major attention on the orbit's roles in the phase curves.

\subsection{Composite velocity fields}
%
The composite 
kinematics of clouds in the observer's frame is the supervision of virial motion and orbital motion.
Calculations of the velocity fields are complicated but straightforward, involving the transformation 
of projected velocity of clouds located at different locations \citep[see details in][]{Wang2018}. 
In the observer's frame, one cloud at point $P$ in the $O-X_{1}Y_{1}Z_{1}$ plane 
is moving with virial velocity ($\bm{V}_{\rm vir}$) (i.e., the Keplerian velocity) around its host black 
hole and supplement with the orbital motion of $\bm{V}_{\rm orb}$. We thus have the projected velocity 
of the cloud of the BLR-{\it i} in the observer's frame
\begin{equation}
V_{{\rm obs},i}=\bmn\vdot \left(\bm{V}_{{\rm vir},i}+\bm{V}_{{\rm orb},i}\right),
\end{equation}
where $\bm{V}_{{\rm orb},i}=\bm{\Omega}\times \bm{R}_{i}$, $\bm{R}_{i}=\bm{\calA}_{i}+\bm{r}_{i}$ is 
the distance vector from the mass center, $\bm{\calA}_{i}$ is the vector of $OO_{i}$ line, 
$\bm{V}_{{\rm vir},i}=\bm{\omega}_{i}\times \bm{r}_{i}$,
and $|\bm{\omega}_{i}|=\sqrt{GM_{i}/r_{i}^{3}}$ is the angular velocity of BLR-{\it i}. 
We express the projected velocity explicitly as
\begin{align}\label{eq:vobs1}
V_{\rm obs,1} = &\,\Omega[-\calA_{1}\cos\phi_0 + r_1\cos(\phi_{0}+\phi_{1})\cos\phi_{\rm B1} 
- r_1\cos\theta_1\sin(\phi_{0}+\phi_{1})\sin\phi_{\rm B1}]\sin i_0 \nonumber \\
&+ V_{\rm K,1}\left\{ \left[-\cos i_{0} \sin\theta_{1} + \cos\theta_1 \sin i_{0} \cos(\phi_{0}+
\phi_{1})\right]\cos\phi_{\rm B1} -\sin i_{0} \sin(\phi_{0}+\phi_{1})\sin\phi_{\rm B1}\right\},
\end{align}
for BLR-I, and 
\begin{align}\label{eq:vobs2}
V_{\rm obs,2} = &\, \Omega[\calA_{2}\cos\phi_0 - r_2\cos(\phi_{0}+\phi_{2})\cos\phi_{\rm B2} 
+ r_2\cos\theta_2\sin(\phi_{0}+\phi_{2})\sin\phi_{\rm B2}]\sin i_0 \nonumber \\
&+ V_{\rm K,2}\left\{ \left[-\cos i_{0} \sin\theta_{2} - \cos\theta_2 \sin i_{0} \cos(\phi_{0}
+\phi_{2})\right]\cos\phi_{\rm B2} + \sin i_{0} \sin(\phi_{0}+\phi_{2})\sin\phi_{\rm B2}\right\},
\end{align}
for BLR-II, where $V_{{\rm K},i}=|\bm{V}_{{\rm vir},i}|$ and $\calA_{1}=(1-\mu_{1})|\bm{\calA}_{0}|$
and $\calA_{2}=\mu_{1}|\bm{\calA}_{0}|$ are the distances of the two black holes to the mass center, 
$\theta_{1,2}$ and $\phi_{1,2}$ are the polar angle and azimuth angle of the normal line of cloud's 
orbital plane, respectively, $\phi_{\rm B1,B2}$ is the orbital phase of the cloud, and $r_{1,2}$ is the 
distance of the cloud to its BH.

Let's see roles of some parameters in the projected velocity. 
The first term of the right side of Equation (\ref{eq:vobs1}) and (\ref{eq:vobs2}) is due to orbital
motion of the binary. This term relates with separations and radius, but also with
orbital phase angles and inclinations. The second term of the rightside of Equation
(\ref{eq:vobs1}) and (\ref{eq:vobs2}) presents the virial motion, but influenced by $i,\phi_{0}$ and
$\theta_{1}$ (by $\Theta$ after integrating over the BLRs).
We would like to emphasize the different roles of the orbital and virial motions in the profiles of
broad emission lines and the differential phase curves. The first mainly drives shifts of the central
wavelength of the lines, giving rise to asymmetry of profiles
and phase curves. The second broadens emission lines as well as the phase curves. 
Though the resultant dependence of profiles and phase curves are mixed, their properties can be
understood separately for a given binary system.

\begin{figure}
\centering
\includegraphics[width=0.95\textwidth]{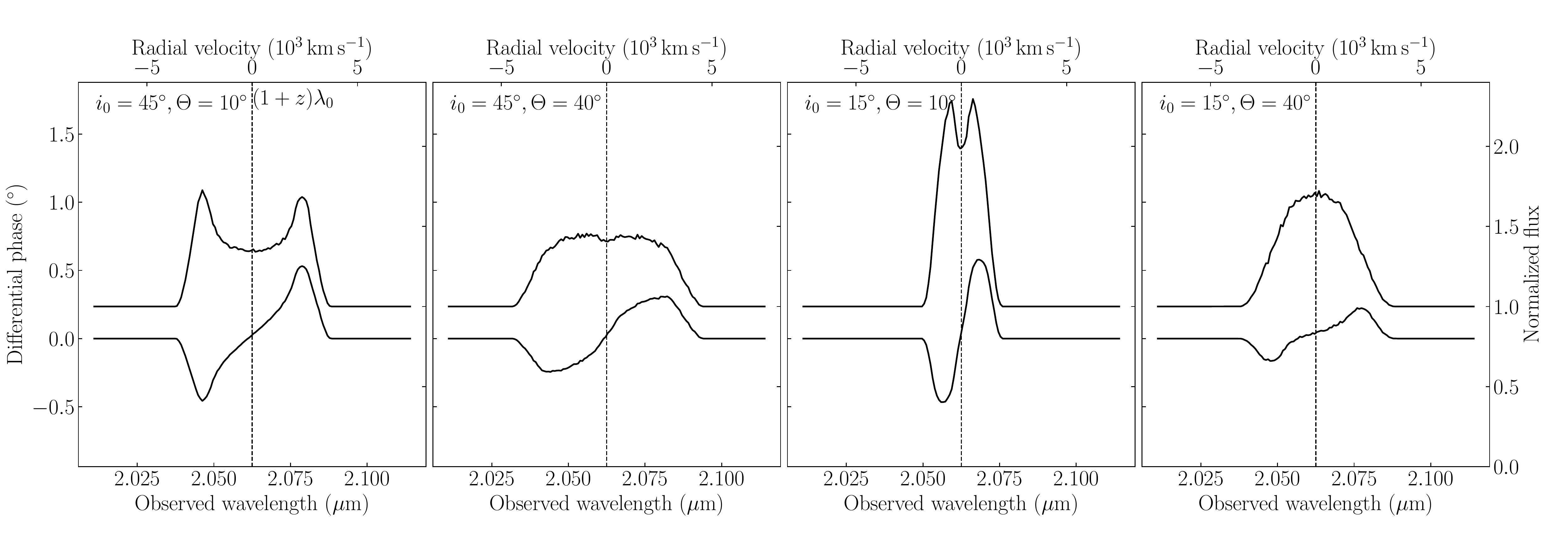}
\figcaption{\footnotesize Line profiles (upper line) of a singe disk BLR 
and differential phase curves for different inclinations and thickness. The differential
phase curves are regarded as the $S$-shape changing with inclinations and thickness of the BLR,
but the $S$-shape alway appear whatever the line profile is double or single peaked. 
We assume $10^{5}$ clouds in the BLR. The dashed line is wavelength of $(1+z)\lambda_{0}$,
which is used in other plots.}
\label{fig:single}
\end{figure}

Here we assume that the orbital motion additional to each BLR is approximated by rigid body
since the orbital period of the CB-SMBH is longer than the dynamical timescale of each BLR. 
This can be simply justified by the orbital period of the CB-SMBH is longer
than the dynamical timescale of each BLR, namely, 
${\cal T}_{\rm obr}\gtrsim \left(R_{\rm BLR}^{3}/GM_{\bullet}\right)^{1/2}$ because
$A_{0}\gtrsim R_{\rm BLR}$ generally holds for the detached binary BLRs.
Otherwise, the BLR will be twisted by the orbital motion, making
profiles complicated as well as the phase curves.

\section{A single BLR}
Table \ref{tab:parameters} lists all the parameters used in the present model.
The normal $R-L$ relation is employed to set up the BLR of AGNs \citep{Bentz2013,Du2018}. We use 
$L_{5100}=\lambda_{\rm Edd}L_{\rm Edd}/\kappa_{\rm bol}=
1.26\times 10^{44}\,\kappa_{10}^{-1}\lambda_{0.1}M_{8}$\,erg\,s$^{-1}$ 
for the typical 5100\AA\, luminosity from accreting black holes, where
$\kappa_{10}=\kappa_{\rm bol}/10$ is the bolometric correction factor, 
$\lambda_{0.1}=\lambda_{\rm Edd}/0.1$ and $\lambda_{\rm Edd}=L_{\rm Bol}/L_{\rm Edd}$
is the Eddington ratio, $L_{\rm Bol}$ is bolometric luminosity 
and $M_{8}=\bhm/10^{8}\sunm$ is black hole mass. We then get the BLR size of 
$R_{\rm BLR}\approx 38\,\kappa_{10}^{-1/2}\lambda_{0.1}^{1/2}M_{8}^{1/2}$\,ltd for each BLR
from the $R-L$ relation \citep{Bentz2013}. Assuming binary black holes have a same Eddington ratio, 
the luminosity ratio is reduced to black hole mass ratio. We keep the total mass of the binary 
system $M_{\rm tot}=10^{8}\sunm$ for all the calculations. The separations are varied in Figure 
4{\it c} and 4{\it d}.  In order to conveniently compare with 
results from GRAVITY, we take the 
typical baseline as $B=100\,$m. We also assume that the baseline is chosen to be perpendicular to
the rotation axis (PA\,$=90^{\circ}$) so that the baseline is along the $\lambda$-photoncentre 
displacement for pure Keplerian case ($\bm{u}\parallel \bm{\epsilon}_{\rm BLR}$) \citep{Rakshit2015}.
The emission line for optical interferometry is Pa-$\alpha$ line 
(its rest wavelength is $\lambda_{0}=1.875\mu$m) 
from targets with a redshift of $z=0.1$ (giving an angular size distance of around $380\rm Mpc$). 
The equivalent width of the Pa-$\alpha$ line is fixed at $0.02\,\rm \mu m$ for the case that the 
line is strong enough ($f_{\ell}\sim 0.3$). 

In our calculations, we adjust ($\Theta$, $i_{0}$, $\phi_0$, $\calA_{0}$, $\mu_{1}$) for their 
effects on the phase curves in 
order to compare the results with the characteristics of CB-SMBH. Other parameters are fixed at 
typical values. Figure 2 shows various profiles of single BLR for different inclinations and thickness. 
It is well understood, for a single Keplerian circular disk, that profiles change from a single to a 
double-peaked shape with increases of inclinations. Moreover, the larger opening 
angles of BLR (i.e., thickening the BLR), the higher dispersion of projected velocity to the LOS. 
As results, sharp peaks of the profiles are smeared and the profiles are broadened. 
Actually line profiles are driven jointly by virial motion of clouds and the BLR inclinations. 

We also calculate differential phase curves
of single BLRs accordingly in Figure 2. The curves are known as the $S$-shape. There are
a peak ($\lambda_{1},\Phi_{1}$) and a valley ($\lambda_{2},\Phi_{2}$) appearing symmetrically 
around the central wavelength [$(1+z)\lambda_0$]. The separations between
the peak and the valley wavelength ($\Delta \lambda_{12}=|\lambda_{1}-\lambda_{2}|$) 
are jointly governed by inclinations for a given BLR. Full-width-half maximu of the peak
and the valley ($\Delta\lambda_{1},\Delta\lambda_{2}$) are determined by inclinations and thickness 
of the BLR (i.e., the dispersion velocity of clouds in the BLR). The amplitudes of the peak and
the valley $\left(|\Phi_{1}|,|\Phi_{2}|\right)$ are controlled by size and thickness of BLR-disks. 
Amplitudes $|\Phi_{1,2}|$ are getting smaller for larger $\Theta$.
This is caused by the shifts of the photoncentre towards the black hole in the baseline
direction when increasing the thickness.
This BLR size linearly increases $\Phi_{1}$ and $\Phi_{2}$. We omit plots about this.

Here we would like to stress the assumption that all clouds have ordered virial motion. 
\cite{Sturm2018} invoked this assumption in the GRAVITY-based measurements of the BLR size and
black hole mass, however, as pointed out by \cite{Stern2015}, the phase curves
should be sensitive to the distribution of angular momentum of clouds in the BLR.  
In principle, the distribution is determined by the formation of the BLR \citep[e.g.,][]{Wang2017a}. 
It is then anticipated to reveal formation of the BLR if the GRAVITY is able to measure the angular
momentum distribution of clouds. For simplicity, we assume all clouds of individual BLRs
have either clockwise or anti-clockwise rotation in this paper, but it is definitely worth of
exploring this dependence qualitatively in a future paper.

\section{Binary BLRs}
Before investigating binary BLRs, we digress for a moment to examine 
the case of binary stars. If thermal motion and star size
can be neglected compared with orbital motion of the binary and the separation, line profile 
is composed of two $\delta$-functions
with central wavelengths of $\lambda_{\pm}=(1\pm v/c)\lambda_{0}$, respectively, where 
$v$ is the orbital velocity and $c$ is the speed of light, and the peak ratio is determined 
by the luminosity ratio of each star (or the binary mass ratio). The corresponding differential 
phase curve is then proportional to $\delta(\lambda-\lambda_{+})-q\delta(\lambda-\lambda_{-})$, 
where $q$ is ratio related with the binary system. Between $\lambda_{+}$ and $\lambda_{-}$,
the phase is zero as a plateau. This is an extreme case of binary BLRs when clouds have zero
virial velocity. The differences from the normal binary stars are the non-negligible sizes 
of each BLR and their virial motion compared with orbital separation and orbital rotating 
velocity, respectively. The two factors change the velocity fields projected to the 
line-of-sight field.

\begin{table}
\footnotesize
\centering
\caption{Parameters in three BLR models \label{tab:parameters}}
\begin{tabular}{llccc}\hline\hline
	Parameters &Meanings  & BLR-I & BLR-II  & fiducial values\\ \hline
    $M_{\bullet}(10^{7}{\sunm})$ & BH mass & $M_1$ & $M_2$ & ($6,4$) \\
    $r_{\rm in}$(ltd) & inner radius of BLR & $r_{\rm in}^{\rm I}$ & $r_{\rm in}^{\rm II}$& (20,16)\\
    $r_{\rm out}$(ltd) & outer radius of BLR & $r_{\rm out}^{\rm I}$ & $r_{\rm out}^{\rm II}$& (40, 32)\\
    $\lambda_{\rm Edd}$& Eddington ratios    &$\lambda_{\rm Edd}^{\rm I}$&$\lambda_{\rm Edd}^{\rm II}$& (0.1,0.1)   \\
    $\gamma$ & power index of emissivity & $\gamma_{1}$ & $\gamma_{2}$ & (2.0,2.0)  \\ 
    $\rm EW(\mu m)$ & equivalent width of the emission line & ${\rm EW_1}$ & $\rm EW_{2}$& ($0.02, 0.02$) \\
    $\Theta$ & thickness of each BLR     & $\Theta_{1}$ & $\Theta_{2}$ & ($10^{\circ},10^{\circ}$) \\ 
    $\bmell$ & angular momentum          &    $\bmell_{1}$  &  $\bmell_{2}$ & (+,+)\\ 
    $\xi_{0}$& reprocessing coefficient at $r_{\rm in}$& $\xi_{0,1}$ & $\xi_{0,2}$& (1.0,1.0)\\ \hline
    $\bm{\calA}_{0}$(ltd) & separation between binary black holes& 100\\
    $\bmell_{\rm orb}$   & orbital angle momentum of the binary  & +\\
    $\phi_{0}$         & orbital phase angle                &    $0^{\circ}$   \\
    $i_{0}$                & inclination angle of the binary system   & $45^{\circ}$ \\
    \hline
\end{tabular}
\tablecomments{\footnotesize Here we assume circle orbit of the binary system in this paper.
Fiducial numbers follow the sequence of the BLR-I and -II.} 
\end{table}

\begin{sidewaysfigure}
\includegraphics[width=1.0\columnwidth]{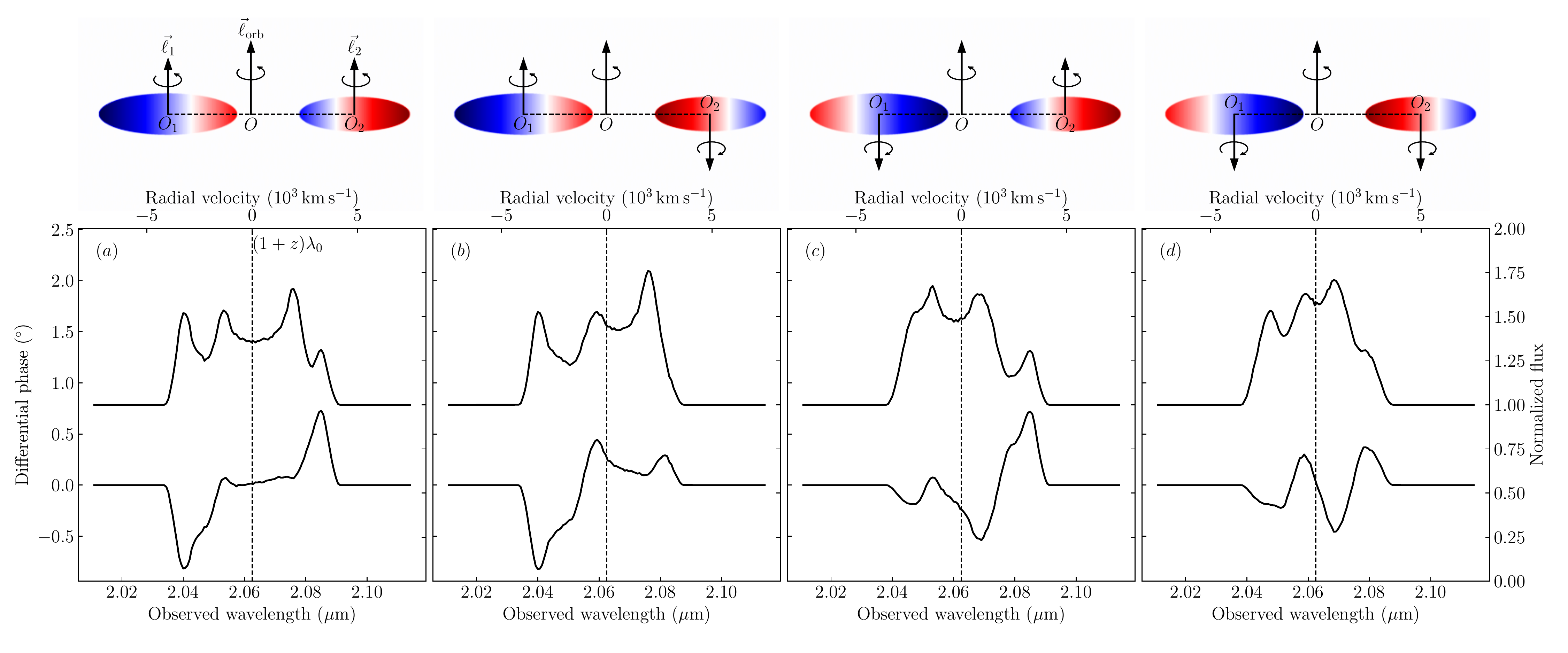}
\caption{\footnotesize Line profiles of a binary thin-disk BLRs ($\Theta=10^{\circ}$) 
and differential phase curves of four simplest groups in light of combinations of angular momentum
(Case-I, -II, -III and -IV). 
{\it Upper panels}: the distributions of projected velocity fields are changed by the orbital motion
in the observer's frame. Approaching clouds are emitting blue-shifted
Pa-$\alpha$ photons (colored as blue regions) whereas receding clouds are emitting red-shifted 
photons (as red regions). The color gradients are determined by composition of virial motion of 
clouds and orbital motions. Deeper color 
regions are due to enhancement of virial and orbital motions whereas shallower regions are due to 
cancellation of the two kinds of motions.
{\it Lower panels}: the simple $S$-shaped phase curves (in Figure 2) are seriously modified by the 
binary system. Composition of the virial motion of clouds and orbital motion changes photoncentres 
due to Doppler shifts, resulting in changes of the phase curves. Profiles of emission line and their
phase curves are attached below the cartoons of four groups of angular momentum compositions.
Panel {\it a}: a plateau appears between the peak and the valley located at ($\lambda_{1},\phi_{1}$) 
and ($\lambda_{2},\phi_{2}$), respectively, we denote it as ``the phase plateau''.
Panel {\it b}: the phase plateau is replaced by a new peak in the slightly blue side.
Panel {\it c}: the phase plateau is replaced by a new peak and a new valley.
Panel {\it d}: a new pair of peak and valley replaces the phase plateau on the both sides of 
the central wavelength of the emission line. See text for details of results.}
\label{fig:AM}
\end{sidewaysfigure}

In principle, the differential phase curves are jointly controlled by two major  
aspects: 1) projected velocities of cloud's motion; and 2) spatial distributions of 
clouds in the space concerned. In the context of binary BLRs, the composed velocity of
the local motion of clouds
bounded by their host SMBH and the orbital motion of the binary is the intrinsic factor 
controlling the projected velocity (inclinations and orbital phase angles are external factors).  
Moreover, the composed velocity is sensitive to the relative directions of angular momentum 
of each BLR and orbital motion resulting in complicated distribution of the projected velocity 
to the LOS. In an homogeneous disk of BLR, the approaching (in blue) and receding (in red) 
parts are symmetric to the observer. Interestingly, the orbital motion additional to the 
virial motion of clouds obviously breaks the symmetry. The upper panels of Figure \ref{fig:AM} 
draw colored regions in blue and in red with gradients from a shallow level to the deep, showing 
the inhomogeneous distribution of the projected 
velocity. Deeper blue regions (DBRs) indicate approaching to the observer with higher projected
velocity whereas deeper red regions (DRRs) have faster receding velocity. It is clear that both
DBRs and DRRs are formed due to enhancement of velocity composition of cloud's virial and binary 
orbital motion whereas the shallower blue regions (SBRs) and shallower red regions (SRRs) are 
due to cancellation of the composition. The complex composition results in asymmetric distributions
of projected velocity and totally change the photoncentre for a given wavelength, giving rise to
modifications of the phase curves. It is then expected that the phase curves deliver information
of orbital motion of the binary system.

Beside cloud distribution in individual BLR, separations of the binary system play a role
in spatial distributions of clouds as well as photoncentres. For a given binary,
separations related with total mass and periods lead to dependence of the phase curves on
them, but this monotonically increases amplitudes of differential phases.
We will illustrate that the differential phase curves show very distinguished features as indicators of
CB-SMBH for a promising way of hunting them.

\subsection{Effects of angular momentum}\label{momentum}
In order to show the impact of angular momentum effects on profiles and phase curves, we have
to distinguish three angular momentum:  two individual BLRs denoted as
$\bmell_{1}$ and $\bmell_{2}$, and the orbital as $\bm\ell_{\rm orb}$. Here we stress
that $\bmell_{1}$ and $\bmell_{2}$ are the total angular momentum of all clouds in each BLR. 
We know little of the formation of a single BLR \citep[e.g.,][]{Wang2017a}, and
much less of the binary BLRs. It is assumed in this paper that the three
angular momentum are independent of each other, so there are four simplest groups of the binary 
systems according to their relative directions: 
1) Case-I: ($\bmell_{1},\bmell_{\rm orb},\bmell_{2})=(\uparrow,\uparrow,\uparrow)$;
2) Case-II: ($\bmell_{1},\bmell_{\rm orb},\bmell_{2})=(\uparrow,\uparrow,\downarrow)$, 
3) Case-III: ($\bmell_{1},\bmell_{\rm orb},\bmell_{2})=(\downarrow,\uparrow,\uparrow)$, and
4) Case-IV: ($\bmell_{1},\bmell_{\rm orb},\bmell_{2},)=(\downarrow,\uparrow,\downarrow)$ for 
binary BLRs. Here the symbol of $\uparrow$ means upward direction of angular momentum,
and $\downarrow$ down direction. These are shown in Figure \ref{fig:AM}. All the four cases 
are assumed to be co-planed, namely, they are either 
parallel or anti-parallel. Cases with non-parallel directions remain for future discussions.

For Case-I, we calculate phase curves for wide ranges of parameters of the binary model. We 
consider CB-SMBHs with parameters of $(i_0,\phi_{0},\Theta_{1},\Theta_{2},\mu_{1},\calA_{0})=
(45^{\circ},0^{\circ},10^{\circ},10^{\circ},0.6,100{\rm ltd})$ as a reference 
to illustrate its complicated dependence on parameters. The line profile and its corresponding 
phase curve of this reference are shown in Figure \ref{fig:AM}{\it a}.
Doppler shifts of each BLR driven by the orbital motion split the entire profile into several 
peaks, giving rise to appearance of triple or even quadruple peaks. 
Comparing with Figure 2{\it a}, the peak and the valley of the phase curve
of the binary BLRs remain similar because both the DBRs and DRRs are located at the
both sides outward the binary system in Case-I. However, the 
part between the peak and the valley is totally
changed, showing a plateau in Figure 3{\it a}. We denote this feature as ``a phase plateau''
which never appears in that of a single
BLR in various ranges of parameters. We regard it as a unique feature of binary BLRs in Case-I.
The plateau is formed through cancellation of phases of photons produced by BLR clouds between 
the two disk-BLRs, where the composite velocities of clouds are getting vanished due to the virial
and orbital motion giving rise to $\lambda$-photoncentre's shift accordingly and hence differential 
phase. The width of the phase plateau is determined by the projected orbital velocity. Amplitudes 
of the peak and the valley are governed by distances of the clouds in the DBRs and the DRRs
to the mass center.
The wavelengths of the peak and the valley are jointly controlled by the projected velocity 
of orbital motion and the virial motion of clouds.

Figure 3{\it b} shows profiles and phase curves of Case-II, where
the receding BLR-II has the anti-direction ($\bmell_{2}$) with the orbital, namely, 
$\bmell_{2}\vdot\bmell_{\rm orb}=-\ell_{2}\ell_{\rm orb}<0$. 
The DBRs remain there, but the DRRs shift to mass center of the binary and
decrease the maximum of red phase peak. In the meanwhile, the SBRs shift 
to the end outward from the inward, causing a small peak in the blue side of the phase curve
and replacing the plateau. This is shown in Figure 3{\it b}. 
The phase plateau is broken by the down-direction of BLR-II angular momentum.
The $\lambda_2$-peak is greatly reduced, the $\lambda_1$-valley remains.

We show results of Case-III in Figure 3{\it c}. BLR-I is approaching to the observer with
a down-direction angular momentum. Comparing with Case-I, the DBR and the SRR of BLR-I 
exchange their location, but composited velocity field of BLR-II remains.
The $\lambda_{1}$-valley is greatly modified into two small valleys, but $\lambda_2$-peak 
remains. The plateau is changed. Case-III has a similar phase curve to Case-II, but the 
red and blue part of phase curve exchange.

Case-IV is shown in Figure 3{\it d}, where the two BLRs have parallel angular momentum, but
just opposite to the orbital. This leads to shifts of (DBRs, DRRs) and (SBRs,SRRs) to their
opposite locations. The values of $\lambda_1$ valley are greatly reduced whereas the SBRs
form a small peak. In the meanwhile, the SRRs shift to end outward, giving rise to a new 
deep valley whereas the DRRs shift to end inward and formed a reduced red peak. The plateau
is replaced by a new peak and new valley in Case-IV.
It should be noted that the binary BLRs with $\mu_{1}=0.6$ showing something asymmetric modified 
by the composition of the virial and orbital motion.  

Finally, if the three angular momentum have fully random directions, both line profiles and 
phase curves have more complicated shapes, but are among the four extreme cases discussed here.
Calculations are extremely tedious and will be carried out separately.

\begin{figure}
\centering
\includegraphics[width=0.67\textwidth]{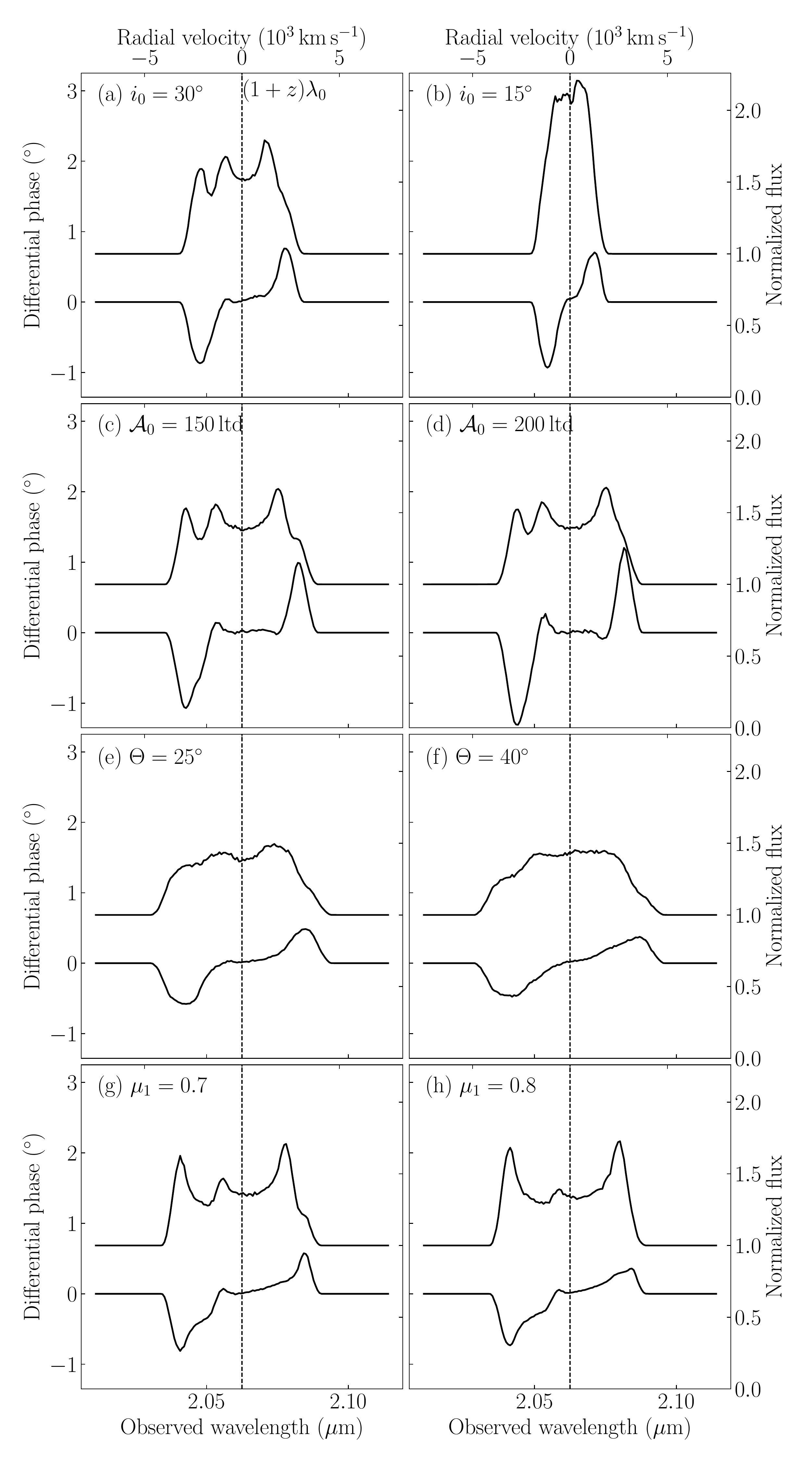}
\figcaption{\footnotesize Effects of four parameters ($i_0,\calA_0,\Theta,\mu_1$) on line profiles and
differential phase curves. We take $(i_0,\calA_{0},\Theta,\mu_{1})=(45^{\circ},100\,{\rm ltd},10^{\circ},0.6)$
as typical values of CB-SMBHs. One parameter with its values is indicated in each panel for its roles 
in the $S$-shape curves. }
\end{figure}

\begin{figure}
\centering
\includegraphics[width=1.02\textwidth]{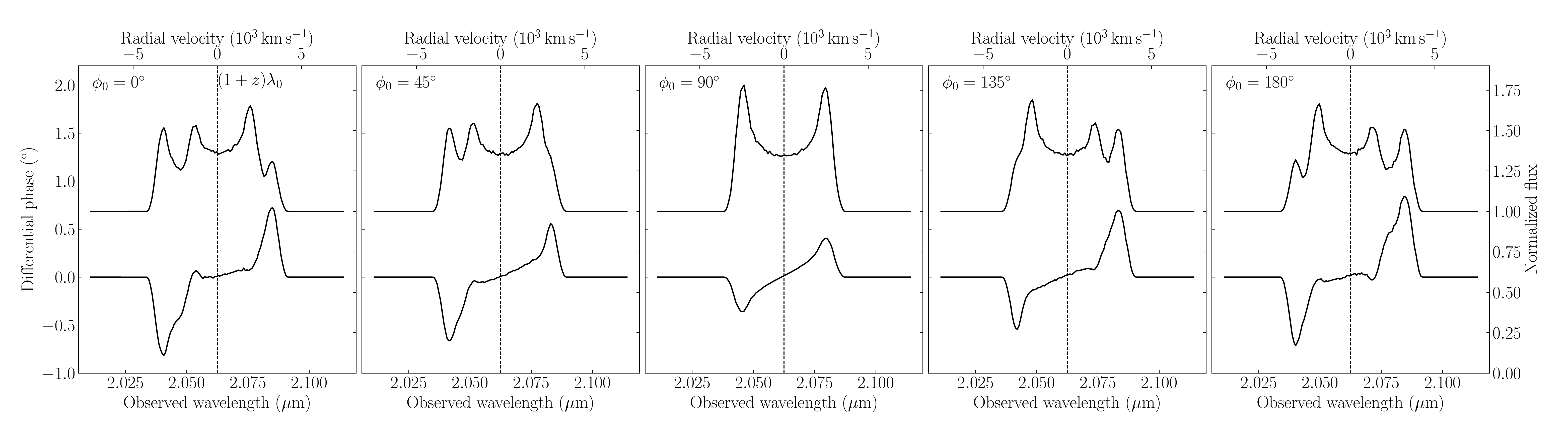}
\figcaption{\footnotesize Series of the differential phase curves with orbital phases from
$\phi_{0}=0^{\circ}$ to $180^{\circ}$. Anti-symmetry of the phase curves appear when 
$\phi_{0}^{\prime}=\phi_{0}+90^{\circ}$. This provides additional evidence for orbital motion
of CB-SMBHs for future observations.
}
\end{figure}

\subsection{Dependence of phase curves on parameters}
Given the angular momentum, the $S$-shape of the phase curves will be modified by the following 
factors: 1) inclinations;
2) orbital phase angles; 3) half opening angles; 4) separations of the two SMBHs and 
5) orbital velocity of the CB-SMBH (depending on mass ratio, total mass and the separations).
Given parameters of binary BLRs in Figure 3{\it a}, we illustrate the roles of each parameter 
of the Case-I binary BLRs in phase curves in Figures 4 and 5.  Line profile and corresponding 
phase curve are compared to show the roles of each parameter. 

Figure 3{\it a} and 
(4{\it a}, 4{\it b}) show that the phase plateau
width decreases with $i_0$ from high inclinations ($i_0=45^{\circ}$) to low. It will disappear 
for pure face-on binary BLRs. The values of peak and valley phases slightly change,
but their wavelengths shift toward the $(1+z)\lambda_{0}$ dramatically. 
Line profiles gets narrower with inclinations.

{\cblue As shown by Figure 4({\it c,d}),the amplitudes of both the peak and the valley get 
larger linearly as shown by Figure 4({\it c,d)} when increasing the separations of the binary 
BLRs. This can be easily understood by the increases of path difference of photons with 
separations.}
 
Increasing the thickness of each BLR results in broadening of line profiles, also the $S$-shaped
phase curves. This can be found by comparing Figure 3{\it a} with Figure 4({\it e, f}). However,
the plateau widths decrease with $\Theta$. The amplitudes, ($\Phi_{1},\Phi_{2}$) 
are intermediately sensitive to $\Theta$, but $(\lambda_{1},\lambda_{2}$) insensitive. 
The phase plateau is still visible even for $\Theta=40^{\circ}$ (which is very close to that 
in 3C 273), but it could be indistinguishable for more thicker disk BLRs.
 
Mass ratio of the binary system represents the relative contribution of each component to the total.
Figure 4({\it g,h}) show the dependence of profiles and phase curves on $\mu_{1}$. Asymmetry of 
line profiles decrease with increases of $\mu_{1}$, but that of phase curves increase with $\mu_{1}$
obviously. The slop of the phase plateau changes with $\mu_1$ obviously, the plateau disappears 
when $\mu_1\rightarrow 1$ (decreasing the secondary BLR leads to lower the phase signals from it). 

Projected velocity of orbital motion changes with orbital angles, but the virial motion of clouds
in each BLR remain the same. From Figure 5, we find 
that amplitudes of $\Phi_1$ and $\Phi_{2}$ sensitively
change with $\phi_0$, but $\lambda_{1}$ and $\lambda_{2}$ only slightly do with $\phi_0$.
The phase plateau changes its slop with $\phi_{0}$. When $\phi_{0}=90^{\circ}$, the projected
velocity of the orbital motion vanishes so that profiles and phase curves reduce to the case similar
to a single BLR as shown in Figure 2{\it a}. In such a case, both ($\Phi_{1},\Phi_{2}$) decrease 
simultaneously. {\cblue We calculate a series of the phase curves for phase angles in Figure 5.
It shows additional evidence for orbital motion of CB-SMBHs if such a series is found over 
a reasonable length of monitoring campaigns. }

Finally we would like to point out the roles of the binary parameters in Case-II, -III and -IV.
Given angular momentum of the binary system, $(i_0,\Theta,\phi_{0},\mu_{1},\calA_{0})$ are playing similar 
roles in Case-II, Case-III and Case-IV similar to Case-I, such as
on amplitudes of peaks and valleys, broadening of profiles and phase curves.

\subsection{Combination of different geometries}
If the binary BLRs are composed of one thin-disk and one thick-disk, the differential phase
curves are a kind of combination of Figures 3{\it a} and 4{\it f}. 
A binary BLRs composed of an approaching thin-disk 
and a receding thick-disk BLRs have a pair of narrower blue peaks and a pair of broaden 
red peaks in profiles, hence,
a narrower blue valley and broaden red peak appear in the phase curve. In principle, properties of
any combination can be understood, and we thus omit plots of these contexts. We limit the present 
paper for disk-like BLRs, however, it is also possible that the CB-SMBHs have combinations among 
(disk, inflows, outflows), even mixture at some levels. Moreover, without justifications from
observations, we assume two BLRs and orbital planes are co-planed for simplicity. If two BLRs 
and orbital planes are fully random, the phase cures must be modified also. 
These situations will share something interesting, 
but different from the present. These are beyond the scope of this paper.

In a brief summary, we show the differential interferometric signatures of close binaries 
of supermassive black holes. It has been found that the well-known 
$S$-shaped curves of differential phases are complicated by two main factors:
orbital motion and the BLR size. The reddest peak and bluest valley (e.g., Case-I) always 
appear, but their amplitudes change with angular momentum of the two BLRs and their orbit. 
The part between them either have a plateau, or multiple peaks and valleys accordingly. The 
curves are also sensitive to the parameters of binary BLRs at different levels, such as 
thickness and orientation of the BLR. It is obvious that orbital information 
of binary BLRs will be delivered by the observed phase curves.

\section{Discussion}
\subsection{Future identification}
The phase curves calculated in this paper shed light on the future identifications of CB-SMBHs.
As we shown in Figures \ref{fig:single} and \ref{fig:AM}, binary BLRs usually have complicated
profiles. However, targets with multiple peaked profiles do not necessarily mean the presence
of CB-SMBH because inhomogeneous BLR has complicated profiles too.   
Actually, observational identifications of CB-SMBHs are so hard that non has been done for sure.
Monitoring AGNs with H$\beta$ Asymmetry (MAHA) campaign focuses on this population of AGNs to
search for CB-SMBH \citep{Du2018,Brotherton2019}, and some of them (with declinations lower 
than $20^{\circ}$) are bright enough for the GRAVITY. Future targets for the GRAVITY
can be selected from most plausible candidates monitored by the MAHA. 
Moreover, fortunately, reverberation mapping technique measures delays of broad emission line 
with respect to the varying continuum \citep{Blandford1982}, and hence is independent of 
measurements of the BLR from the different phase curves. In particular, the velocity-resolved
delay maps are able to distinguish kinematics and geometry of the BLR \citep{Peterson2014} and the
2-D transfer functions deliver information of orbital motion of CB-SMBHs \citep{Wang2018}. 
{\cblue It usually takes a few hours for the GRAVITY to perform observations of an individual target 
whereas RM campaigns do a few months 
or even a few years. Considering potential variations of BLRs with a timescale of
$\Delta t_{\rm BLR}\sim c\tau_{\rm BLR}/V_{\rm FWHM}=3.4\,\tau_{20}V_{5000}^{-1}\,$yrs, 
RM-campaigns should be done around
the epoch of GRAVITY observations in order to reduce the difference of BLRs measured by the two
different ways. Here $\tau_{20}$ is the H$\beta$ lags in units of 20days and 
$V_{5000}=V_{\rm FWHM}/5000{\rm km\,s^{-1}}$ is the FWHM of the H$\beta$ profile. 
We call this quasi-simultaneous epoch (QSE) observations.
Joint analysis of the QSE observations} will generate complete information of the 
BLR including angular momentum distributions of BLR clouds and diagnostic evidence for CB-SMBHs.

{\cblue An excellent targets is Ark 120 ($z=0.0328$, $K\approx 10$mag, 
${\rm RA=05h\,16m\,11.42s,\,Dec=-00d\,08m\,59.4s}$) 
from the MAHA project \citep{Du2018}, which are of asymmetric and double-peaked
profiles of broad H$\beta$ line. Moreover, Ark 120 has long-term variations with a period 
of about 20\,yrs \citep{Li2019}. The GRAVITY will detect Brackett $\gamma$ line 
($\lambda_{0}=2.166\mu$m). The semimajor axis of the binary’s orbit is about 27.0 ltd 
in Ark 120 if it is a binary \citep{Li2019}. Low-cadence RM-campaign shows that
H$\beta$ lags are about 40days during 1995-1997 \citep{Peterson1998}, 
but the MAHA observations show a lag of $\sim 16\,$days during 2016-2017 \citep{Du2018}. 
We have conducted 
a campaign with a cadence of about 3days since 2015 and it shows interesting features
of H$\beta$ reverberation. It would be greatly interesting for the GRAVITY to explore 
the central regions of Ark 120 and a joint analysis of RM observations is highly desired.
}

The present model generally works for identification of CB-SMBHs from candidates 
since ${\cal T}_{\rm orb}\gtrsim \tau_{\rm BLR}$ holds, where
$\tau_{\rm BLR}\sim R_{\rm BLR}/V_{\rm FWHM}$ is the dynamical timescales of BLRs. 
This guarantees the necessary condition that binary BLRs hold during the orbital period. 
Astronomers may make multiple measurements of GRAVITY to test changes of phase 
curves with orbital motion. This also can yield orbital parameters of the CB-SMBH if the phase curves are
measured accurately enough by GRAVITY or next generation OASIS \citep{Abuter2017}. This will allow astronomers
to compare orbital parameters with the properties of low-frequency gravitational waves.

\subsection{Orbital parameters of CB-SMBH}\label{sec:para}
As listed in Table 1, there are 18 independent parameters in one CB-SMBH. It should
be pointed out that ($\phi_0,\Theta,\mu_1$) are degenerate somehow in modifying the $S$-shapes 
of the phase curves. This degeneration leads to some difficulties in measuring the binary parameters
from the phase curves. Actually the phase curves provide spatial information of
the regions perpendicular to the LOS. Fortunately, reverberation mapping measure the response of 
clouds along the LOS direction and provides another way of measuring black hole masses 
through the Chain Monte Carlo (MCMC) simulations \citep{Pancoast2011, Li2013, Li2018}. 
Accuracy of SMBH mass can reach $\sim 40\%$ in Mrk 142 \citep{Li2018}.
Jointly fitting the phase curves and RM light curves along with the
changes of profiles will provide the most reliable measurements of one CB-SMBH (Songsheng et al.
2019 in preparation).

The characterized frequency of GWs from the typical CB-SMBHs is 
\begin{equation}
f_{\rm GW}=\frac{2}{{\cal T}_{\rm orb}}=0.3\, M_{8}^{1/2} \calA_{100}^{-3/2} \, {\rm nHz}.
\end{equation}
Unlike
the $10^{2}$Hz GWs detected by the LIGO, low-frequency GWs keep the constant waveform making it 
hard to observe chirping waveforms before the final states of mergers. Though it has been suggested
to measure the orbital parameters through the PTA technique \citep{Sesana2010}, degeneracies of
parameters of the binary system make it hard to test the properties of the low-frequency GWs.
This strongly suggests necessaries of independent determinations of orbital parameters of the 
CB-SMBHs to break some degeneracies, for example,
joint observations of the GRAVITY and reverberation mapping campaigns. This provides an excellent
comparison to examine results obtained by the PTA measurements.

\subsection{CB-SMBH with merged BLR}
The present model of binary BLRs still has two detached BLRs. When merger proceeds,
the secondary is spiraling into the primary BLR, leading to redistribution of clouds of 
the binary BLRs. Merged consequences of Case-I, -II, -III and -IV will be different as
results of $(\bmell_1+\bmell_2)$-composition because the orbital angular momentum is 
efficiently radiated away by GWs
in the four cases. In Case-I and -IV, the resultant angular momentum of merged BLR probably 
remain the specific angular momentum of the BLR prior to the merger keeping a virialized 
disk BLR (but the two cases could be slightly different due to the orbital angular momentum). 
However the resultant angular momentum is getting to
vanish in Case-II and -III (i.e., $|\bmell_{1}+\bmell_{2}|\ll |\bmell_{1}|+\bmell_{2}|$),
which will generate very different results of phase curves from Figure 3{\it a}.
The real situations of ongoing mergers are much complicated than that of the present 
considerations, it is important to note these consequences. 
It would be interesting to find the merged BLRs from GRAVITY observations.
Additionally, the effects of tidal influence on the BLRs will 
distort regular geometry of BLR discussed here during the merging process.
This subject remains open.

{\cblue
We note there is a stage that each SMBHs could have their own torus (or outposts
partially merged). In such a case, the binary SMBHs could be spatially resolved by 
MATISSE (the Multi AperTure mid-Infrared SpectroScopic Experiment). The evolution 
of the binary is still governed by dynamical friction rather than gravitational waves.
}

\subsection{Alternative of binary BLRs}\label{sec:alter}
Asymmetric
profiles as an indicator of binary black holes are a long-term debate to identify CB-SMBH.
Alternatively, the asymmetry can be explained by precessing spiral arms in disk 
\citep{Eracleous1995, Storchi-Bergmann2003}, or hot spot \citep{Newman1997,Jovanovic2010}. 
In these models, however, the hot spot or spiral arm is only 
photoionized by the same one ionizing source. This naturally results in the same reverberation 
of the blue and red parts with with respect to ionizing power. Actually, the campaign of 
Fairall 9 with a strong red asymmetry of H$\beta$ line show the same reverberation properties
of the blue and red \citep{Santos-Lleo1997}. Moreover, non-axisymmetrical disk-like BLR, such 
as elliptical 
disks also lead to asymmetric profiles of broad emission lines as well as the asymmetric
curves of differential phase. Inhomogeneous components as sub-structures can be done by 
a useful tool of MICA (Multiple Inhomogeneous Component Analysis) \citep{Li2016b}.
All of these cases can be justified by reverberation mapping campaign. 
Analysis of 3C 390.3 and 3C 120 from the MAHA project will be carried out separately.

For binary BLRs photoionized by their own iozning sources,
fortunately, broad emission lines should have much more complicated response with respect to
the varying continuum than that of other models (i.e, the hot spot, the spiral arm or elliptical
orbiting clouds). 2D-transfer functions are conveniently obtained by campaigns
of reverberation mapping of AGNs for this goal \citep{Wang2018}.
This lends opportunity to distinguish CB-SMBHs from AGNs.

\section{Concluding remarks}
Motivated by the unprecedented high spatial resolution of the Very Large Telescope Interferometry 
(VLTI) with successful application to the BLR of 3C 273, we explore the differential interferometric 
signature of close binary of supermassive black holes. In principle, the differential phase curves
are determined by the velocity fields projected to the line of sight. The well-known $S$-shaped
differential phase curves of a single BLR will be modified by the orbital motion in the binaries.
This provide an easy way of identifying the binaries through GRAVITY.
The phase curves of the binaries are sensitive to
\begin{itemize}
\item combinations of the angular momentum of the two BLRs and the orbit, showing different
characteristics of the phase curves with the phase plateau, multiple
peaks and valleys. These characteristics as indicators of orbital motions relative to the BLR 
virial motion is delivering information of the binary system. 
 
\item Except for dependence on the parameters of the binary system, the phase curves also 
strongly depend on parameters of observer's parameters, such as, inclinations and orbital 
angles. Roles of parameters in the phase curves degenerate at some level, but could be relaxed
by joint observations of reverberation mapping campaigns measuring kinematics of ionized gas.

\item The present paper only illustrates the characterized results of the simplest situations. 
Properties of general contexts will share something intermediate among the present results. 
All the detailed results of other geometries and random angular momentum will be carried out 
separately.

\end{itemize}

As a summary of the present paper, the current GRAVITY
is able to detect ``twisted" $S$-shaped curves as a consequence of orbital motion of close
binaries of supermassive black holes. We expect for the GRAVITY to spatially resolve them  
in active galactic nuclei in near future. As an independent measurement,
reverberation mapping of broad emission lines probing properties of the same regions, in principle, 
is a tool supplementary to the interferometry of spatially resolving the center around the SMBHs,
not only for BH mass, but also for detection of binary black holes. {\cblue Joint analysis of the 
GRAVITY and reverberation mapping observations becomes an indispensable
tool to test low-frequency gravitational waves detected by PTAs.}

\vglue 0.5cm
We thank a anonymous referee for a useful report to clarify a few points.
We acknowledge the support by National Key R\&D Program of China (grants 2016YFA0400701 and
2016YFA0400702), by NSFC through grants {NSFC-11873048, -11833008, -11573026, -11473002, -11721303, 
-11773029, -11833008, -11690024}, and by Grant No. QYZDJ-SSW-SLH007 from the Key Research Program
of Frontier Sciences, CAS, by the Strategic Priority Research Program of the Chinese Academy of 
Sciences grant No.XDB23010400.

\clearpage
\appendix
\section{Coordinates of BLR-I and BLR-II}
The displacement of BLR-I to the mass center is given by $\bm{\calA}_{1}=(1-\mu_{1})\bm{\calA}_{0}$. 
We have the projected displacement to the observer 
\begin{equation}
\bm{\calA}_{\rm obs}=(1-\mu_{1})(\bmn\vdot\bm{\calA}_{0})\bmn, 
\end{equation}
and the displacement on the tangent plane of the sky is given by
\begin{equation}
\bm{\calA}^{\prime}=(1-\mu_{1})\bm{\calA}_{0}-(1-\mu_{1})(\bmn\vdot\bm{\calA}_{0})\bmn
                   =(1-\mu_{1})\left[\bm{\calA}_{0}-(\bmn\vdot\bm{\calA}_{0})\bmn\right]
\end{equation}
and
\begin{equation}
\bm{\alpha}^{\rm c}_{1}=\frac{\bm{\calA}_{0}^{\prime}}{D_{\rm A}}
                     =\frac{\bm{\calA}_{0}-(\bmn\vdot\bm{\calA}_{0})\bmn}{D_{\rm A}}\,(1-\mu_{1}).
\end{equation}
Similarly, we have coordinates of BLR-II
\begin{equation}
\bm{\alpha}^{\rm c}_{2}=-\frac{\bm{\calA}_{0}-(\bmn\vdot\bm{\calA}_{0})\bmn}{D_{\rm A}}\,\mu_{1}.
\end{equation}

\begin{figure}
\centering
\includegraphics[width=0.5\textwidth]{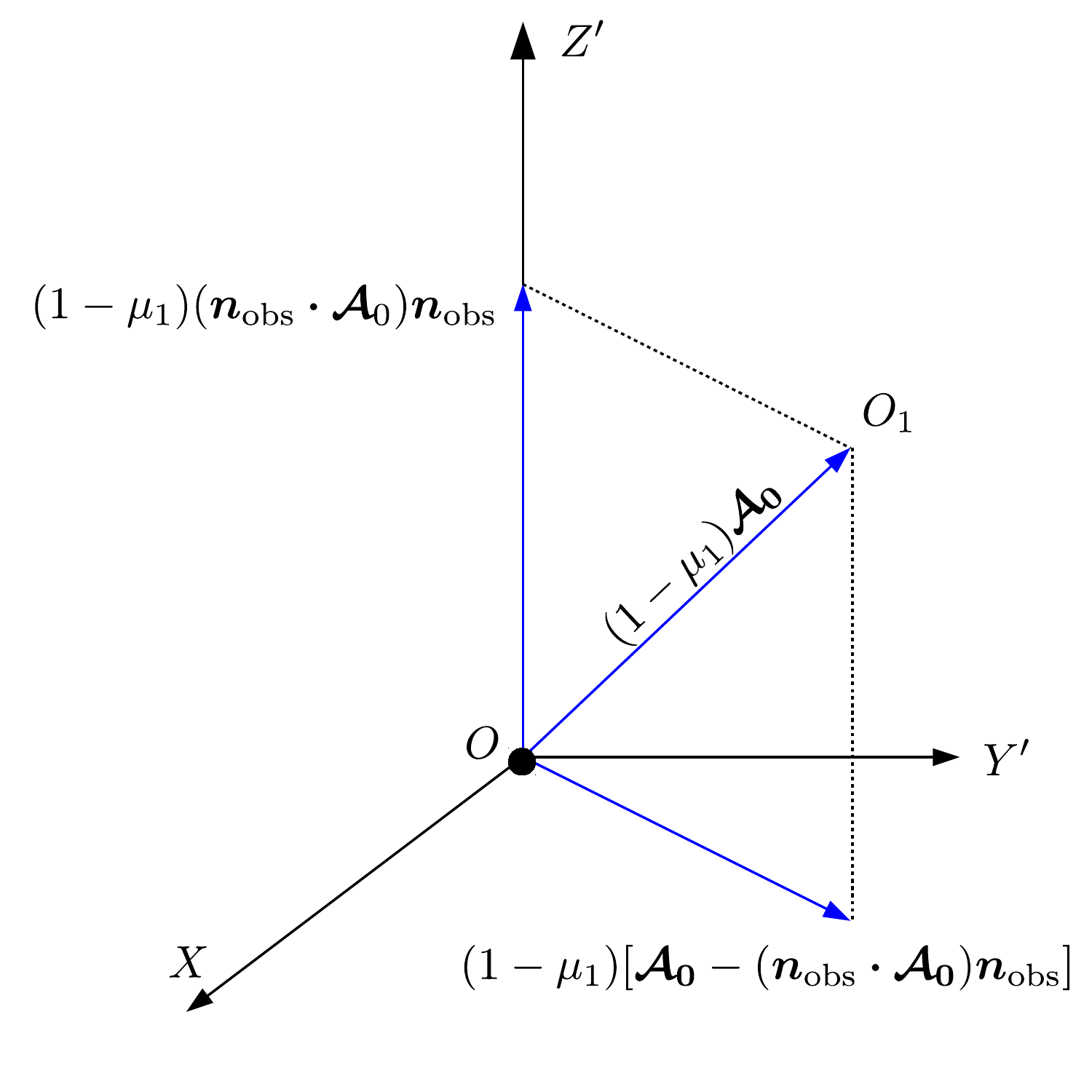}
\figcaption{\footnotesize Coordinates of two individual BLRs are given in
the tangent plane of the sky. The symbols have the same meanings with Figure 1, but the 
orientations are plotted for convenience here. }
\end{figure}

\end{document}